\documentclass[twoside,a4paper,11pt]{sea10}
\pdfoutput=1
\usepackage{graphicx}
\usepackage{hyperref}
\usepackage{movie15}
\begin{document}
\pagenumbering{arabic}
\pagestyle{myheadings}
\thispagestyle{empty}
{\flushleft\includegraphics[width=\textwidth,bb=58 650 590 680]{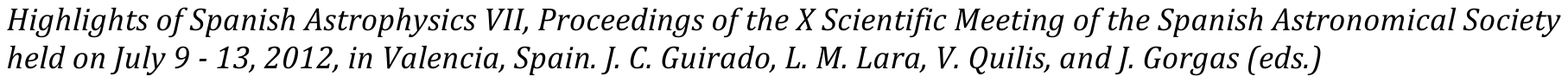}}
\vspace*{0.2cm}
\begin{flushleft}
{\bf {\LARGE
%
Status of the OTELO Project
%
}\\
\vspace*{1cm}
%
J. Cepa$^{1,2}$,
A. Bongiovanni$^{1}$, 
A.M. P\'erez Garc\'\i a$^{1}$, 
E.J. Alfaro$^{3}$,
H.O. Casta\~neda$^{4}$,
A. Ederoclite$^{1}$,
J.J. Gonz\'alez$^{5}$,
J.I. Gonz\'alez--Serrano$^{6}$,
M. S\'anchez--Portal$^{7}$,
J. Bland-Hawthorn$^{8}$,
D.H. Jones$^{9}$,
J. Gallego$^{10}$,
and 
J.M. Rodr\'\i guez--Espinosa$^{1}$
%
}\\
\vspace*{0.5cm}
%
$^{1}$
Instituto de Astrof\'\i sica de Canarias, E--38200 La Laguna, Tenerife, Spain\\
$^{2}$
Departamento de Astrof\'\i sica, Universidad de La Laguna, E-38206 La Laguna, Tenerife, Spain\\
$^{3}$
Instituto de Astrof\'\i sica de Andaluc\'\i a (CSIC), Apdo. 3004,  E--18080, Granada, Spain \\
$^{4}$
Departamento de F\'\i sica, Escuela Superior de F\'\i sica y
Matem\'atica, IPN, M\'exico D.F., Mexico\\
$^{5}$
Instituto de Astronom\'\i a, Universidad Nacional Aut\'onoma de
M\'exico, Apdo Postal 70-264, 
Cd. Universitaria, 04510, Mexico\\
$^{6}$
Instituto de F\'\i sica de Cantabria, CSIC-Universidad de Cantabria, Santander, Spain\\
$^{7}$
Herschel Science Centre, ESAC/INSA, P.O. Box 78, 28691 Villanueva
de la Ca\~nada, Madrid\\
$^{8}$
Sydney Institute for Astronomy, School of Physics, University of Sydney, NSW 2006, Australia\\
$^{9}$
Anglo-Australian Observatory, PO Box 296, Epping, NSW 1710, Australia\\
$^{10}$
Departamento de Astrof\'\i sica, Universidad Complutense de Madrid, 28040 Madrid, Spain\\
%
\end{flushleft}
%
\markboth{
The OTELO Project
}{ 
%
Cepa et al.
%
}
\thispagestyle{empty}
\vspace*{0.4cm}
\begin{minipage}[l]{0.09\textwidth}
\ 
\end{minipage}
\begin{minipage}[r]{0.9\textwidth}
\vspace{1cm}
\section*{Abstract}{\small
%
The OTELO project is the extragalactic survey currently under way 
using the tunable filters of the
OSIRIS instrument at the GTC. OTELO is already providing the deepest 
emission line object survey of the 
universe up to a redshift 7. In this contribution, the status of the survey 
and the first results obtained will be presented.
%
\normalsize}
\end{minipage}
%
%
%
\section{Introduction \label{intro}}
The study of the evolution of galaxies and of high redshift objects has undergone a substantial advance in the last decade due to the deep multi--wavelength extragalactic surveys available. These surveys could be classified according to the spectral resolution. Low resolution broad--band multicolour photometry allows observing the faintest targets, and deriving redshift estimations together with morphological parameters. Mid--band multicolour surveys allow increasing redshift accuracy, and even detecting the brigthest line emitters, with penalty on depth. Higher resolution spectroscopic surveys allow obtaining accurate redshifts, and more detailed spectral energy distributions for tackling a wide variety of scientific objectives. Intermediate resolution narrow band imaging surveys, such as OTELO \cite{cepa08}, are a powerful tool to detect and study the evolution of line emitter objects (see Steidel et al. \cite{steidel00} and references 
therein). They allow, depending on the emission line observed according to the redshift of the source, deriving star 
formation rates (SFR), metallicities and its cosmic evolution, for every target in the field without previous selection. Then, narrow band surveys complement broad band surveys, that are more efficient in detecting continuum dominated and bright emission line targets, and complement spectroscopic surveys, whose targets are selected using broad band surveys. In summary, narrow band surveys 
provide a complementary view of the universe at high redshift.

Within narrow band surveys, those using tunable filters as OTELO, CADIS (Calar Alto Deep Imaging Survey, 
Thommes et al. \cite{thommes97}), and the TTFFGS (Taurus Tunable Filter Faint Galaxy Survey, 
Jones \& Bland-Hawthorn \cite{jones01}) 
detect one order of magnitude more objects (normalizing for telescope size and exposure time) than
conventional narrow band surveys as the Suprime-Cam of Subaru (Fujita et al. \cite{fujita03}). For this reason 
narrow band 
surveys with Tunable Filters in large telescopes constitute a deep sky probe with unprecedented sensitivity. Moreover, since tunable filter surveys obtain a set of images of the same pointing at slightly different wavelengths, this technique can be rather considered narrow band 3D wide field spectroscopy than conventional narrow band imaging. They allow obtaining redshifts with an almost spectroscopic accuracy, while allowing a quite precise photometric calibration for deriving absolute SFRs.

For these reasons, OSIRIS provides GTC with unique capabilities compared with similar telescopes, and the OSIRIS Tunable Emission Line Object survey (OTELO) will supply a unique database in survey area, sensitivity, redshift accuracy and target discrimination, as shown in Table 1.

\section{OTELO Survey}

OTELO \cite{cepa08} is aimed at surveying emission line objects using OSIRIS
tunable filters in selected atmospheric windows relatively free of sky emission lines. Different high latitude and
low extinction sky regions with enough angular separations will be observed yielding a total area of 0.1 square
degrees. A minimum detectable flux of 5$\times 10^{-19}$ erg/cm$^2$/s will
allow detecting objects of equivalent width (EW) of 3\AA\ or smaller, 
making OTELO the deepest emission line survey to date (Table 1). This 
lowest EW will allow detecting, for the first time in this kind of
surveys, even faint spirals 
and blue compact dwarf galaxies at redshifts up to 1.5. OTELO is a deep space probe that will provide a representative 
sample of the Universe from z =0.4 through 7.0. Given the observing procedure, OTELO will allow studying clearly
defined volumes of Universe at a known flux limit.

To this aim, 108 dark hours of guaranteed observing time at a single pointing at the Extended Groth Strip (EGS) will be devoted for obtaining 
images of 36 contiguous wavelengths at a FWHM of 1.2nm, scanning every 0.6nm (i.e.: half the FWHM) in the 
907--928nm window 
in the OH sky line forrest. Each wavelength will be observed 6600 seconds distributed in 6 exposures of 1100 
seconds dithered 18 arcseconds in a cross--shaped pattern to fill out the gap between detectors.

\begin{table}[ht] 
\caption{OTELO Survey main characteristics} 
\center
\begin{minipage}{0.5\textwidth}
\center
\begin{tabular}{cc} 
\hline\hline 
Parameter & Value \\ [0.5ex]   
\hline 
Limiting flux (3$\sigma$) & 5$\times$10$^{-19}$erg/cm$^2$/s\\ 
Minimum EW & 3\AA\\
Area & 0.1 sq.deg.\\
Redshift accuracy & 10$^{-4}$\\
Cosmic statistics & Several fields \\ 
Deblend H$\alpha$ from [NII] & Yes\\ [1ex]  
\hline
\end{tabular} 
\end{minipage}
\label{tab1} 
\end{table}

\begin{figure}
\center
\includegraphics[scale=0.35]{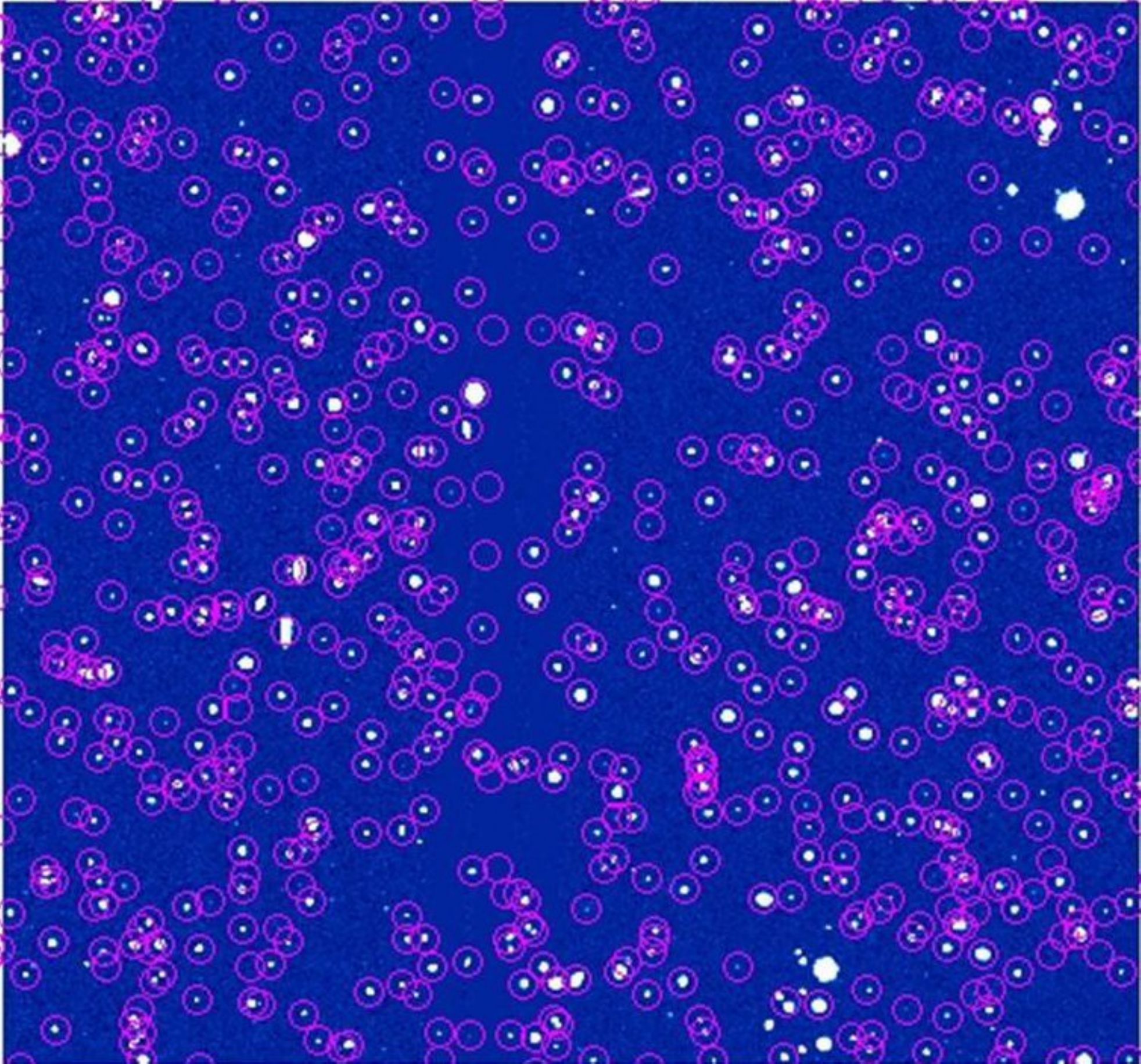}
\caption{\label{fig1} OTELO sources of a 7.4$\times$7.1 arcmin$^2$ section of the 
synthesized NB filter $\sim$924.4/8.4nm, obtained by the median of all matched images from 920.8 through 928.0nm. The sources marked with a circle are those identified in the shallower z$^\prime$ survey of \cite{gwyn11}.
}
\end{figure}

Also, observations of a second pointing in the central part of Lockman Hole, with the same observational configuration but in grey time, will start next season.

\section{OTELO observations and data reduction}

So far 39 hours have been observed corresponding to 13 contiguous wavelengths in the spectral range 
920.8--928.0nm from April 2010 through April 2012. The mean seeing during the observations was of 0.9$\pm$0.2 arcsec, as measured directly 
on the scientific images. The best seeing corresponds to 0.64 arcsec. The TF tuning during the observations 
was found stable at the nominal accuracy of 0.1nm, as expected.

Data reduction was performed using standard IRAF routines. Bias was first subtracted, and the images 
were trimmed. Cosmic rays were removed, and flatfielding was achieved by fitting a 2D surface to the data, 
since sky flats were not available, and dome flats are not useful due to illumination effects. Then, sky rings 
were subtracted using the TFRED IRAF package \cite{jones02} after iteratively masking objects at 3$\sigma$ level. 
A median combination of dithered images provide a fringe map that
was subtracted to the data, providing fringe corrected images. 

For each image, 
astrometry was performed using stars of z$^{\prime}<$23.0
magnitudes of the 
CFHTLS\footnote{CFHTLS is based on observations obtained 
with MegaPrime/MegaCam, a joint project of CFHT and CEA/DAPNIA, at the 
Canada-France-Hawaii Telescope (CFHT) which is operated by the National Research Council (NRC) of Canada, the Institut 
National des Science de l'Univers of the Centre National de la Recherche Scientifique (CNRS) of France, and the University of 
Hawaii. This work is based in part on data products produced at TERAPIX and the Canadian Astronomy Data Centre as part 
of the Canada-France-Hawaii Telescope Legacy Survey, a collaborative project of NRC and CNRS.}, and the matched images of the same wavelength combined (Fig. 2). Finally the sources 
were extracted using SExtractor and the centre to edge wavelength variation corrected based on the TF optical centre 
and the new wavelength dependence derived by the instrument team \cite{gonzalez13}. Flux calibration was achieved using two standard 
spectrophotometric stars within the same field.

\begin{figure}
\center
\includegraphics[scale=0.45]{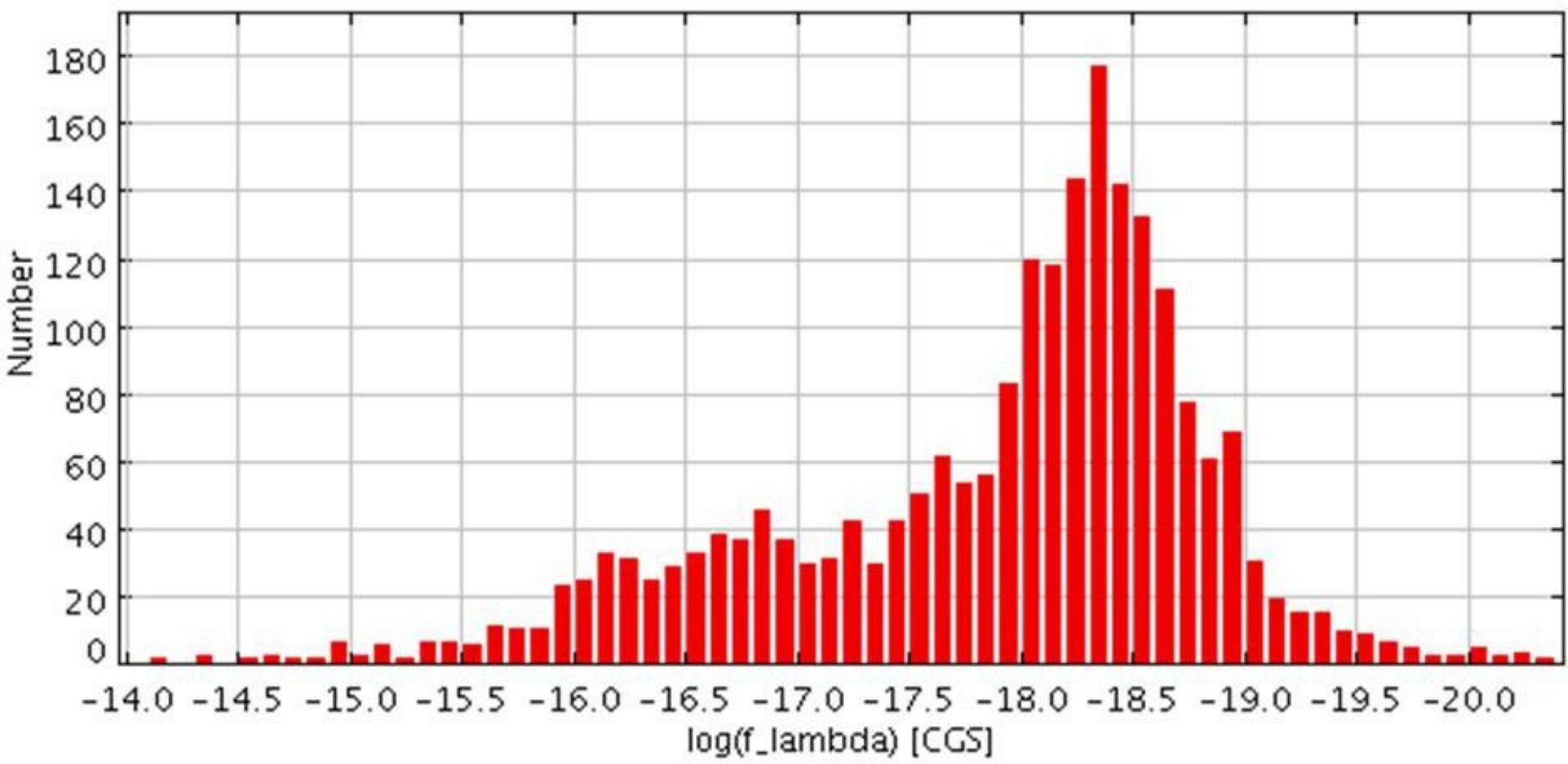}
\caption{\label{fig2} Histogram of the 2300 sources with counterparts in the \cite{gwyn11} catalogue, detected in the image of the synthesized filter 924.4/8.4nm, with an exposure time equivalent to 85800s. A minimum detectable flux of 5$\times 10^{-19}$ erg/cm$^2$/s at 3$\sigma$ level is obtained.
}
\end{figure}

\section{Data quality} 

A median of the 78 OTELO images was obtained, synthesizing a filter 924.4/8.4nm. Then the extracted sources from this median image were cross matched with the catalogue of \cite{gwyn11} obtained using the z$^\prime$ filter, reaching up to magnitude 23rd, shallower than achievable in OTELO survey (Figure 1). A total of more than 2300 sources were identified in this way.

As expected, a minimum detectable flux of 5$\times 10^{-19}$ erg/cm$^2$/s
(3$\sigma$) was achieved, with completitude at 1$\times 10^{-18}$ erg/cm$^2$/s 
(3$\sigma$) (Fig. 2), two magnitudes deeper than the deepest narrow band survey so far available. These figures are fully consistent with the OSIRIS ETC calculators provided by the instrument team and available in the WWW. 

\begin{figure}
\center
\includegraphics[scale=0.5]{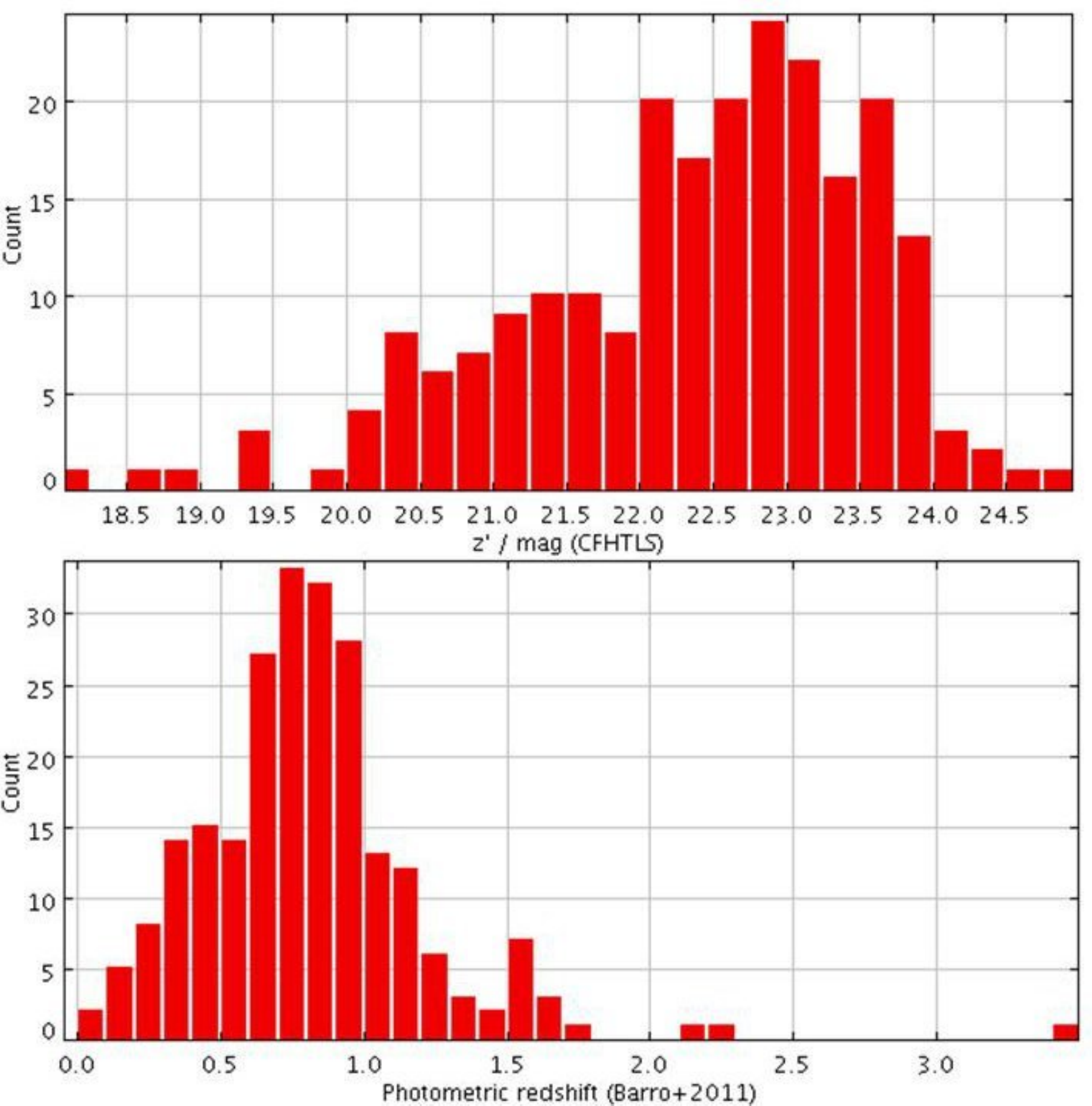}
\caption{\label{fig3} Histogram of the extracted sources with emission line signatures that are detected at all wavelenghts and with counterparts in \cite{barro11}. Top, the distribution of z$\prime$ magnitudes in the shallower survey of \cite{gwyn11}. Down, the photometric redshifts estimated by \cite{barro11}. Peaks corresponding to the most conspicuous lines H$\alpha$, [OIII]$\lambda$500.7nm and [OII]$\lambda\lambda$372.7nm, can be seen at z $\sim$ 0.4, 0.8 and 1.5, respectively. 
}
\end{figure}

\section{Preliminary results}

With one third of the first pointing complete, it is just possible to start deriving some scientific results. As a first step, only sources identified at all wavelengths and with emission line signatures were considered, resulting in a total of 237 emitters. From these, 222 show counterpart in \cite{barro11}. In Figure 3, the histograms of the distribution of z$^\prime$ magnitudes from \cite{gwyn11} and the distribution of the photometric redshifts derived by \cite{barro11} are shown. Most emission line galaxies are detected between redshifts 0.5 and 1.0. As a consequence, the emission line detected is mostly [OIII]$\lambda$500.7nm, with H$\alpha$ at redshifts z $<$ 0.4, and [OII]$\lambda\lambda$372.7nm at z $\sim$1.5 also present. In fact, peaks can be distinguished at redshifts $\sim$ 0.4, 0.8, and 1.5 corresponding to these emission lines (Figure 3). Further analysis of the pseudo--spectra of these sources will allow deriving precise redshifts and line fluxes. 

The next step will be detecting targets present only in two contiguous wavelengths, or even in only one wavelength, corresponding to faint continuum emission line galaxies. For the H$\alpha$ emitters at redshift z $\sim$ 0.4, the [NII] lines will be deblended thus obtaining SFR and a metalicity estimation for these objects \cite{mall}. Finally, matching surveys of the same field in the MIR, FIR and X-ray will allow studying different types of emission line targets. Follow--up optical spectroscopic observations are foreseen. In fact OSIRIS MOS observations of the Lockman Hole field using guaranteed time of the instrument team are already planned, and are expected to be executed as soon as this mode is available.

\section{Summary}
 
Once completed, OTELO will be a unique survey in terms of minimum detectable flux
and EW limit, yielding the 
deepest emission line survey to date with spectroscopic redshift accuracy. In this contribution one third of data of the first pointing are analyzed. Data gathering and analysis will continue during 2013 for EGS and Lockman Hole fields.

%
%
\small  
%
\section*{Acknowledgments}   
%

This work was supported by the Spanish Ministry of Economy and Competitiveness (MINECO) under the grant AYA2011--29517--C03--01.
Based on observations made with the Gran Telescopio Canarias (GTC), installed in the Spanish Observatorio del Roque de los 
Muchachos of the Instituto de Astrof\'\i sica de Canarias, in the island of La Palma. J.C., E.J.A., A.E., J.G., and J.M.R.E. acknowledge partial support from the Consolider--Ingenio grant CSD00070-2006 (MINECO).
%

%
\end{document}